\batchmode
\makeatletter
\def\input@path{{/home/bercherj/JFB/Recherche/Papers/GSI2013//}}
\makeatother
\documentclass[a4paper,english,resume]{llncs}
\usepackage[T1]{fontenc}
\usepackage[latin9]{inputenc}
\usepackage{amsmath}
\usepackage{amssymb}
\usepackage{esint}

\makeatletter

\pdfpageheight\paperheight
\pdfpagewidth\paperwidth

\makeatother

\usepackage{babel}
\begin{document}

\title{Some results on a $\chi$-divergence, an~extended~Fisher information
and~generalized~Cramér-Rao inequalities
\thanks{To be presented during the conference ``Geometric Sciences of Information'', Paris, August 28-30, 2013 (and a version published in the proceedings)}}
%}
%to be presented during the conference ``Geometric Sciences of Information'', Paris, August 28-30, 2013 (and a version published in the proceedings)}}

\author{Jean-François Bercher}

\institute{Laboratoire d'Informatique Gaspard Monge, UMR 8049 \\ Université
Paris-Est, ESIEE, Cité Descartes 93162 Noisy-le-Grand Cedex \\ \email{jf.bercher@esiee.fr}}
\maketitle
\begin{abstract}
We propose a modified $\chi^{\beta}$-divergence, give some of its
properties, and show that this leads to the definition of a generalized
Fisher information. We give generalized Cramér-Rao inequalities, involving
this Fisher information, an extension of the Fisher information matrix,
and arbitrary norms and power of the estimation error. In the case
of a location parameter, we obtain new characterizations of the generalized
$q$-Gaussians, for instance as the distribution with a given moment
that minimizes the generalized Fisher information. Finally we indicate
how the generalized Fisher information can lead to new uncertainty
relations.  
\end{abstract}

\section{Introduction}

In this communication, we begin with a variation on a $\chi^{\beta}$-divergence,
which introduces an averaging with respect to an arbitrary distribution.
We give some properties of this divergence, including monotonicity
and a data processing inequality. Then, in the context of parameter
estimation, we show that this induces both an extension of the standard
Fisher information, that reduces to the standard Fisher information
in a particular case, and leads to an extension of the Cramér-Rao
inequality for the estimation of a parameter. We show how these results
can be expressed in the multidimensional case, with general norms.
We also show how the classical notion of Fisher information matrix
can be extended in this context, but unfortunately in a non-explicit
form. 

In the case of a translation parameter and using the concept of escort
distributions, the general Cramér-Rao inequality leads to an inequality
for distributions which is saturated by generalized $q$-Gaussian
distributions. These generalized $q$-Gaussians are important in several
areas of physics and mathematics. They are known to maximize the $q$-entropies
subject to a moment constraint. The Cramér-Rao inequality shows that
the generalized $q$-Gaussians also minimize the generalized Fisher
information among distributions with a fixed moment.  In information
theory, the de Bruijn identity links the Fisher information and the
derivative of the entropy. We show that this identity can be extended
to generalized versions of entropy and Fisher information. More precisely,
the generalized Fisher information naturally pops up in the expression
of the derivative of the entropy. Finally, we give an extended version
of the Weyl-Heisenberg uncertainty relation as a consequence of the
generalized multidimensional Cramér-Rao inequality. Due to the lack
of space, we will omit or only sketch the proofs, but will reference
to the literature whenever possible.

\section{Context}

Let $f_{1}(x;\theta)$, $f_{2}(x;\theta)$ and $g(x;\theta)$ be three
probability distributions, with $x\in X\subseteq\mathbb{R}^{k}$ and
$\theta$ a parameter of these densities, $\theta\in\mathbb{R}^{n}$.
We will deal with a measure of a divergence between two probability
distributions, say $f_{1}(x)$ and $f_{2}(x),$ and we will also be
interested in the estimation of a vector $h(\theta)\in\mathbb{R}^{m}$,
with $T(x)$ the corresponding estimator. 

If $\left\Vert .\right\Vert $ is an arbitrary norm on $\mathbb{R}^{m}$,
its dual norm $\left\Vert .\right\Vert _{*}$ is defined by 
\[
\left\Vert Y\right\Vert _{*}=\underset{\left\Vert X\right\Vert \leq1}{\mathrm{sup}}X.Y,
\]
 where $X.Y$ is the standard inner product. In particular, if $\left\Vert .\right\Vert $
is an $L_{p}$ norm, then $\left\Vert .\right\Vert _{*}$ is an $L_{q}$
norm, with $p^{-1}+q^{-1}=1.$ A very basic idea in this work is that
it may be useful to vary the probability density function with respect
to which are computed the expectations. Typically, in the context
of estimation, if the error is $T(x)-h(\theta),$ then the bias can
be evaluated as $B_{f}(h(\theta))=\int_{X}\left(T(x)-h(\theta)\right)\, f(x;\theta)\,\text{d}x=\mathrm{E}_{f}\left[T(x)-h(\theta)\right]$
while a general moment of an arbitrary norm of the error can be computed
with respect to another probability distribution, say $g(x;\theta)$,
as in $\mathrm{E}_{g}\left[\left\Vert T(x)-h(\theta)\right\Vert ^{\beta}\right]=\int_{X}\left\Vert T(x)-h(\theta)\right\Vert ^{\beta}\, g(x;\theta)\,\text{d}x$.
The two distributions $f(x;\theta)$ and $g(x,\theta)$ can be chosen
very arbitrary. However, one can also build $g(x;\theta)$ as a transformation
of $f(x;\theta)$ that highlights, or on the contrary scores out,
some characteristics of $f(x;\theta)$. For instance, $g(x;\theta)$
can be a weighted version of $f(x;\theta)$, i.e. $g(x;\theta)=h(x;\theta)f(x;\theta),$
or a quantized version as $g(x;\theta)=\left[f(x;\theta)\right],$
where $\left[.\right]$ denotes the integer part. Another important
special case is when $g(x;\theta)$ is defined as the escort distribution
of order $q$ of $f(x;\theta)$, where $q$ plays the role of a tuning
parameter: $f(x;\theta)$ and $g(x;\theta)$ are a pair of escort
distributions, which are defined as follows: 

\begin{equation}
f(x;\theta)=\frac{g(x;\theta)^{q}}{\int g(x;\theta)^{q}\text{d}x}\,\,\,\,\text{ and }\,\,\, g(x;\theta)=\frac{f(x;\theta)^{\bar{q}}}{\int f(x;\theta)^{\bar{q}}\text{d}x},\label{eq:PairEscorts-1-1}
\end{equation}
where $q$ is a positive parameter, $\bar{q}=1/q,$ and provided of
course that involved integrals are finite. These escort distributions
are an essential ingredient in the nonextensive thermostatistics context.
Actually, the escort distributions have been introduced as an operational
tool in the context of multifractals, c.f. \cite{chhabra_direct_1989},
\cite{beck_thermodynamics_1993}, with interesting connections with
the standard thermodynamics. Discussion of their geometric properties
can be found in \cite{abe_geometry_2003,ohara_dually_2010}. Escort
distributions also prove useful in source coding, as noticed by \cite{bercher_source_2009}.

Finally, we will see that in our results, the generalized $q$-Gaussians
play an important role. These generalized $q$-Gaussians appear in
statistical physics, where they are the maximum entropy distributions
of the nonextensive thermostatistics \cite{tsallis_introduction_2009}.
The generalized $q$-Gaussian distributions, which reduce to the standard
Gaussian in a particular case, define a versatile family that can
describe problems with compact support as well as problems with heavy
tailed distributions. They are also analytical solutions of actual
physical problems, see e.g. \cite{lutz_anomalous_2003}, \cite{schwaemmle_q-gaussians_2008}
and are sometimes known as Barenblatt-Pattle functions, following
their identification by Barenblatt and Pattle.  We shall also mention
that the Generalized $q$-Gaussian distributions appear in other fields,
namely as the solution of non-linear diffusion equations, or as the
distributions that saturate some sharp inequalities in functional
analysis \cite{del_pino_best_2002}, \cite{cordero-erausquin_mass-transportation_2004}.

\section{The modified $\chi^{\beta}$-divergence}

The results presented in this section build on a beautiful, but unfortunately
overlooked, work by Vajda \cite{vajda_-divergence_1973}. In this
work, Vajda presented and characterized an extension of the Fisher
information popping up as a limit case of a $\chi^{\alpha}$ divergence
(for consistency with our notations in previous papers, we will use
here the superscript $\beta$ instead of $\alpha$ as in Vajda's paper).
Here, we simply make a step beyond on this route. 

The $\chi^{\beta}$-divergence between two probability distributions
$f_{1}$ and $f_{2}$ is defined by 
\begin{equation}
\chi^{\beta}(f_{1},f_{2})=E_{f_{2}}\left[\left|1-\frac{f_{1}}{f_{2}}\right|^{\beta}\right],\label{eq:chidiv}
\end{equation}
with $\beta>1.$ In the case $\beta=2,$ and a parametric density
$f_{\theta}(x)=f(x;\theta),$ it is known that the Fisher information
of $f_{\theta}$ is nothing but 
\[
I_{2,1}[f_{\theta};\theta]=\lim_{|t|\rightarrow0}\,\chi^{2}(f_{\theta+t},f_{\theta})/t^{2}.
\]
In \cite{vajda_-divergence_1973}, Vajda extended this to any $\beta>1,$
and defined a generalized Fisher information as $I_{\beta}[f_{\theta};\theta]=\lim_{|t|\rightarrow0}\,\chi^{\beta}(f_{\theta+t},f_{\theta})/\left|t\right|^{\beta}=E_{f_{\theta}}\left[\left|\frac{\dot{f}_{\theta}}{f_{\theta}}\right|^{\beta}\right]$,
assuming that $f_{\theta}$ is differentiable wrt to $\theta$. Let
us consider again a $\chi^{\beta}$-divergence as in (\ref{eq:chidiv}),
but modified in order to involve a third distribution $g(x;\theta):$
\[
\chi_{g}^{\beta}(f_{1},f_{2})=E_{g}\left[\left|\frac{f_{2}-f_{1}}{g}\right|^{\beta}\right].
\]
Obviously, this formula includes the standard divergence. For $\theta\in\mathbb{R}^{n}$,
let us denote by $\partial_{i}f_{\theta}$ the partial derivative
with respect to the $i^{\mathrm{th}}$ component, and let $t_{i}$
be a vector that increments this component. Then, we have $\lim_{|t_{i}|\rightarrow0}\,\chi^{\beta}(f_{\theta+t_{i}},f_{\theta})/\left|t\right|^{\beta}=E_{g_{\theta}}\left[\left|\frac{\partial_{i}f_{\theta}}{g_{\theta}}\right|^{\beta}\right]$.
Doing this for all components and summing the resulting vector, we
finally arrive at the following definition 
\begin{equation}
I_{\beta}[f_{\theta}|g_{\theta};\theta]=\sum_{i}E_{g_{\theta}}\left[\left|\frac{\partial_{i}f_{\theta}}{g_{\theta}}\right|^{\beta}\right]=E_{g_{\theta}}\left[\left\Vert \frac{\nabla f_{\theta}}{g_{\theta}}\right\Vert _{\beta}^{\beta}\right],\label{eq:gen_fish_g}
\end{equation}
where $\nabla f_{\theta}$ denotes the gradient of $f_{\theta}$ and
$\left\Vert .\right\Vert _{\beta}$ is the $\beta$-norm. A version
involving a general norm instead of the $\beta$-norm is given in
\cite{bercher:hal-00766695}. Vajda's generalized Fisher information
\cite{vajda_-divergence_1973} corresponds to the scalar case and
$g_{\theta}=f_{\theta}$. We will see that this generalized Fisher
information, which includes previous definitions as particular cases,
is involved in a generalized Cramér-Rao inequality for parameter estimation. 

The modified $\chi^{\beta}$-divergence enjoys some important properties:
\begin{property}
The modified $\chi^{\beta}$-divergence has information monotonicity.
This me\-ans that coarse-graining the data leads to a loss of information.
If $\tilde{f_{1},}\tilde{f_{2}}$ and $\tilde{g}$ denote the probability
densities after coarse-graining, then we have $\chi_{g}^{\beta}(f_{1},f_{2})\geq\chi_{\tilde{g}}^{\beta}(\tilde{f_{1}},\tilde{f_{2}})$.
A proof of this result can be obtained following the lines in \cite{amari_alpha_2009}.
A consequence of this result is a data processing inequality: if $Y=\phi(X),$
and if $f_{1}^{\phi},$$f_{2}^{\phi}$ and $g^{\phi}$ denotes the
densities after this transformation, then 
\[
\chi_{g}^{\beta}(f_{1},f_{2})\geq\chi_{g^{\phi}}^{\beta}(f_{1}^{\phi},f_{2}^{\phi}),
\]
 with equality if the transformation $Y=\phi(X)$ is invertible. It
must be mentioned here that this also yields an important data processing
inequality for the generalized Fisher information: $I_{\beta}[f_{\theta}|g_{\theta};\theta]\geq I_{\beta}[f_{\theta}^{\phi}|g_{\theta}^{\phi};\theta]$. 
\end{property}
~
\begin{property}
Matrix Fisher data processing inequality. Consider the quadratic case,
i.e. $\beta=2,$ $\theta\in\mathbb{R}^{n}$ and $t$ an increment
on $\theta$. Assuming that the partial derivatives of $f$ wrt the
components of $\theta$ exist and are absolutely integrable, a Taylor
expansion about $\theta$ gives $f_{\theta+t}=f_{\theta}+\sum_{i}t_{i}\partial_{i}f_{\theta}+\frac{1}{2}\sum_{i}\,\sum_{j}\, t_{i}t_{j}\partial_{ij}^{2}f_{\theta}+\ldots$
Hence, to within the second order terms, 
\[
\chi_{g}^{2}(f_{\theta},f_{\theta+t})=E_{g}\left[\left(\sum_{i}t_{i}\frac{\partial_{i}f_{\theta}}{g}\right)^{2}\right]=E_{g}\left[\sum_{i}\sum_{j}t_{i}t_{j}\frac{\partial_{i}f_{\theta}\partial_{j}f_{\theta}}{g^{2}}\right]=t^{T}I_{2,g}[\theta]\, t,
\]
where $I_{2,g}[\theta]=E_{g}\left[\psi_{g}\psi_{g}^{T}\right]=E_{g}\left[\frac{\nabla f_{\theta}}{g_{\theta}}\frac{\nabla f_{\theta}^{T}}{g_{\theta}}\right]$
is a Fisher information matrix computed wrt to $g_{\theta}$ and $\psi_{g}$
is a generalized score function. By information monotocity, $\chi_{g}^{2}(f_{\theta},f_{\theta+t})\geq\chi_{g^{\phi}}^{2}(f_{\theta}^{\phi},f_{\theta+t}^{\phi})$,
and therefore we get that 
\[
I_{2,g}[\theta]\,\geq I_{2,g^{\phi}}[\theta]
\]
 (the difference of the two matrices is positive semi definite) which
is a data processing inequality for Fisher information matrices. 
\end{property}
~\nopagebreak
\begin{property}
If $T(X)$ is a statistic, then with $\alpha^{-1}+\beta^{-1}=1,$
$\alpha\geq1$, we have 
\begin{equation}
\left|E_{f_{2}}\left[T\right]-E_{f_{1}}\left[T\right]\right|\leq E_{g}\left[\left|T\right|^{\alpha}\right]^{\frac{1}{\alpha}}\,\chi_{g}^{\beta}(f_{1},f_{2})^{\frac{1}{\beta}}.\label{eq:prepa_CR}
\end{equation}
It suffices here to consider $\left|E_{g}\left[T\left(\frac{f_{2}-f_{1}}{g}\right)\right]\right|=\left|E_{f_{2}}\left[T\right]-E_{f_{1}}\left[T\right]\right|$
and then apply the Hölder inequality to the left hand side. 
\end{property}

\section{The generalized Cramér-Rao inequality}

 The Property 3 above can be used to derive a generalized Cramér-Rao
inequality involving the generalized Fisher information (\ref{eq:gen_fish_g}).
Let us consider the scalar case. Set $f_{2}=f_{\theta+t},$ $f_{1}=f_{\theta},$
and denote $\eta=E_{f}[T(X)-h(\theta)].$ Then divide both side of
(\ref{eq:prepa_CR}) by $t$, substitute $T(X)$ by $T(X)-h(\theta),$
and take the limit $t\rightarrow0.$ Assuming that we can exchange
the order of integrations and derivations, and using the definition(\ref{eq:gen_fish_g})
in the scalar case, we obtain
\begin{equation}
E_{g}\left[\left|T(X)-h(\theta)\right|^{\alpha}\right]^{\frac{1}{\alpha}}\, I_{\beta}[f_{\theta}|g_{\theta};\theta]^{\frac{1}{\beta}}\geq\left|\frac{d}{d\theta}\eta\right|\label{eq:CRgineqScalar}
\end{equation}
which reduces to the standard Cramér-Rao inequality in the case $\alpha=\beta=2$
and $g=f$. A multidimensional version involving arbitrary norms can
be obtained by the following steps: (a) evaluate the divergence of
the bias, $\nabla_{\theta}.\, B_{f}(\theta),$ (b) introduce an averaging
with respect to $g$ in the resulting integral (c) apply a version
of the Hölder inequality for arbitrary norms. The proof can be found
in \cite{bercher:hal-00766695} for the direct estimation of the parameter
$\theta$. For the estimation of any function $h(\theta),$ we have
a generalized Cramér-Rao inequality that enables to lower bound a
moment of an arbitrary norm of the estimation error, this moment being
computed wrt any distribution $g$. 
\begin{proposition}
{[}Generalized Cramér-Rao inequality, partially in \cite{bercher:hal-00766695}{]}
Under some regularity conditions, then for any estimator $T(X)$ of
$h(\theta)$, 
\begin{gather}
E_{g}\left[\left\Vert T(X)-h(\theta)\right\Vert ^{\alpha}\right]^{\frac{1}{\alpha}}E_{g}\left[\left\Vert H\,\frac{\nabla_{\theta}f(X;\theta)}{g(X;\theta)}\right\Vert _{*}^{\beta}\right]^{\frac{1}{\beta}}\geq\left|m+\nabla_{h(\theta)}.\, B_{f}(h(\theta))\right|\label{eq:GeneralizedCramerRao-4-1}
\end{gather}
where $\alpha$ and $\beta$ are Hölder conjugate of each other, i.e.
$\alpha^{-1}+\beta^{-1}=1,$ $\alpha\geq1$, with $H_{ij}=\partial\theta_{j}/\partial h(\theta)_{i}$,
and where the second factor in the left side is actually $I_{\beta}[f_{\theta}|g_{\theta};h(\theta)]^{\frac{1}{\beta}}$.

\end{proposition}
Many consequences can be obtained from this general result. For instance,
if one chooses $h(\theta)=\theta$ and an unbiased estimator $T(X)=\hat{\theta}(X)$,
then the inequality above reduces to
\begin{equation}
E_{g}\left[\left\Vert \hat{\theta}(X)-\theta\right\Vert ^{\alpha}\right]^{\frac{1}{\alpha}}E_{g}\left[\left\Vert \frac{\nabla_{\theta}f(X;\theta)}{g(X;\theta)}\right\Vert _{*}^{\beta}\right]^{\frac{1}{\beta}}\geq n.\label{eq:trois}
\end{equation}
Taking $f=g$, we obtain an extension of the standard Cramér-Rao inequality,
featuring a general norm and an arbitrary power; in the scalar case,
we obtain the Barankin-Vajda result \cite{barankin_locally_1949,vajda_-divergence_1973},
see also \cite{weinstein_general_1988}. 

When $h(\theta)$ is scalar valued, we have the following variation
on the theme (proof omitted), which involves a Fisher information
matrix, but unfortunately in a non-explicit form: for any matrix $A$,
we have
\begin{equation}
E_{g}\left[|T(X)-h(\theta)|^{\alpha}\right]^{\frac{1}{\alpha}}\geq\frac{\left|\nabla_{\theta}h(\theta)^{t}\, A\,\nabla_{\theta}h(\theta)\right|}{E_{g}\left[|\nabla h(\theta)^{t}A\,\psi_{g}|^{\beta}\right]^{\frac{1}{\beta}}},\label{eq:version matricielle}
\end{equation}
where $\psi_{g}=\frac{\nabla_{\theta}f(x;\theta)}{g(x;\theta)}$ is
a score function. We define as Fisher information matrix the matrix
$A$ which maximizes the right hand side. In the quadratic case $\alpha=\beta=2,$
and using the inequality $(x^{t}x)^{2}\leq\left(x^{t}Bx\right)\,\left(x^{t}B^{-1}x\right)$
valid for any positive definite matrix $B$, one can check that the
maximum is precisely attained for $A^{-1}=I_{2,g}[\theta]=E_{g}[\psi_{g}\psi_{g}^{t}],$
that is the Fisher information matrix we obtained above in Property
2, in the quadratic case. The inequality reduces to the inequality
$E_{g}\left[|T(X)-h(\theta)|^{2}\right]\geq\nabla_{\theta}h(\theta)^{t}\, A\,\nabla_{\theta}h(\theta)$
which is known in the standard case. 

In the quadratic case, it is also possible to obtain an analog of
the well known result that the covariance of the estimation error
is greater than the inverse of the Fisher information matrix (in the
Löwner sense). The proof follows the lines in \cite[pp. 296-297]{liese_statistical_2008}
and we get that 
\begin{equation}
E_{g}\left[\left(T(X)-h(\theta)\right)\left(T(X)-h(\theta)\right)^{t}\right]\geq\dot{\eta}^{t}I_{2,g}[\theta]^{-1}\dot{\eta}\label{eq:CovVersion}
\end{equation}
 with $\dot{\eta}$ the matrix defined by $\dot{\eta}=\nabla_{\theta}E_{f}[\hat{\theta}^{t}],$ and with equality
iff $\dot{\eta}I_{2,g}[\theta]^{-1}\psi_{g}(X)=\lambda\left(T(x)-h(\theta)\right)$,
with $\lambda>0$.

Let us now consider the case of a location parameter $\theta\in\mathbb{R}^{n}$
for a translation family $f(x;\theta)=f(x-\theta)$. In such a case,
we have $\nabla_{\theta}f(x;\theta)=-\nabla_{x}f(x-\theta)$. Let
us also assume, without loss of generality, that $f(x)$ has zero
mean. In these conditions, the estimator $T(X)=\hat{\theta}(X)=X$
is unbiased. Finally, taking $\theta=0$, the relation (\ref{eq:trois})
leads to
\begin{proposition}
{[}Functional Cramér-Rao inequality{]} For any pair of probability
density functions, and under some technical conditions, 
\begin{equation}
\left(\int_{X}\left\Vert x\right\Vert ^{\alpha}g(x)\text{\,\ d}x\right)^{\frac{1}{\alpha}}\left(\int_{X}\left\Vert \frac{\nabla_{x}f(x)}{g(x)}\right\Vert _{*}^{\beta}g(x)\text{\,\ d}x\right)^{\frac{1}{\beta}}\geq n,\label{eq:CRInequalityLocation}
\end{equation}
with equality if  $\nabla_{x}f(x)=-K\, g(x)\|x\|^{\alpha-1}\nabla_{x}\|x\|.$
\end{proposition}
At this point, we can obtain a interesting new characterization of
the generalized $q$-Gaussian distributions, which involves the generalized
Fisher information. Indeed, if we take $f(x)$ and $g(x)$ as a pair
of escort distributions, with $\bar{q}=1/q,$ 

\begin{equation}
f(x)=\frac{g(x)^{q}}{\int g(x)^{q}\text{d}x}\,\,\,\,\text{ and }\,\,\, g(x)=\frac{f(x)^{\bar{q}}}{\int f(x)^{\bar{q}}\text{d}x},\label{eq:PairEscorts-1}
\end{equation}

\begin{proposition}
{[}$q$-Cramér-Rao inequality \cite{bercher:hal-00766695}{]} For
any probability density $g,$ if $m_{\alpha}[g]=E\left[\left\Vert x\right\Vert ^{\alpha}\right]$
is the moment of order $\alpha$ of the norm of $x,$ and if 
\begin{equation}
I_{\beta,q}\left[g\right]=\left(q/M_{q}\left[g\right]\right)^{\beta}\, E_{g}\left[g(x)^{\beta(q-1)}\left\Vert \nabla_{x}\ln g(x)\right\Vert _{*}^{\beta}\right]\label{eq:Ig}
\end{equation}
is the Fisher information of order ($\beta,q$), with $M_{q}[g]=\int g(x)^{q}\text{d}x,$
then 
\begin{equation}
m_{\alpha}[g]^{\frac{1}{\alpha}}\, I_{\beta,q}\left[g\right]^{\frac{1}{\beta}}\geq n\label{eq:ExtendedqCRloc1}
\end{equation}
with equality if and only if $g(x)$ is a generalized Gaussian of
the form 
\begin{equation}
g(x)\propto\left(1-\gamma(q-1)\|x\|^{\alpha}\right)_{+}^{\frac{1}{q-1}}\label{eq:gaussgen}
\end{equation}

\end{proposition}
Let us simply note here that $g(x)$ is also called stretched $q$-Gaussian,
become a stretched Gaussian for $q=1,$ and a standard Gaussian when
in addition $\alpha=2.$ The inequality (\ref{eq:ExtendedqCRloc1})
shows that the generalized $q$-Gaussians minimize the generalized
Fisher information among all distributions with a given moment. 
 Let us also note and mention that the inequality (\ref{eq:ExtendedqCRloc1})
is similar, but different, to an inequality given by Lutwak et al
\cite{lutwak_extensions_2012} which is also saturated by the generalized
Gaussians (\ref{eq:gaussgen}). Finally, for a location parameter,
the matrix inequality (\ref{eq:CovVersion}) reduces to $E_{g}\left[XX^{t}\right]\geq(I_{2,g})^{-1}=E_{g}\left[\psi_{g}\psi_{g}^{t}\right]^{-1},$
with equality iff $f$ is a generalized $q$-Gaussian with covariance
matrix $(I_{2,g})^{-1}$. \\

The generalized Fisher information (\ref{eq:Ig}) also pops up in
an extension of the de Bruijn identity. This identity is usually shown
for the solutions of a heat equation. An extension is obtained by
considering the solutions of a doubly-nonlinear equation \cite{vazquez_smoothing_2006}
\begin{equation}
\frac{\partial}{\partial t}f=\Delta_{\beta}f^{m}=\text{div}\left(|\nabla f^{m}|^{\beta-2}\,\nabla f^{m}\right).\label{eq:dnle}
\end{equation}

\begin{proposition}
{[}Extended de Bruijn identity \cite{bercher:hal-00766699}{]} For
$q=m+1-\frac{\alpha}{\beta}$, $M_{q}[f]=\int f^{q}$ and $S_{q}[f]=\frac{1}{1-q}\left(M_{q}[f]-1\right)$
the Tsallis entropy, we have 
\begin{align}
\frac{\text{d}}{\text{d}t}S_{q}[f] & =\left(\frac{m}{q}\right)^{\beta-1}M_{q}[f]^{\beta}\, I_{\beta,q}[f].\label{eq:ExtendedDeBruijnb}
\end{align}

\end{proposition}
Of course, the standard de Bruijn identity is recovered in the particular
case $\alpha=\beta=2,$ and $q=m=1.$

We close this paper by indicating that it is possible to exhibit new
uncertainty relations, beginning with the generalized Cramér-Rao inequality
(\ref{eq:ExtendedqCRloc1}). These inequalities involve moments computed
with respect to escort distributions like (\ref{eq:PairEscorts-1}).
We denote by $E_{q}[.]$ an expectation computed wrt an escort of
order $q.$ If $\psi$ is a wave function, $x$ and $\xi$ two Fourier
conjugated variables, then 
\begin{proposition}
{[}Uncertainty relations {]} For $k=\beta/\left(\beta(q-1)+1\right),$
$\lambda=n(q-1)+1$, and $\gamma\geq2,$ $\theta\geq2$, 
\begin{alignat}{1}
 & \,\frac{M_{\frac{k}{2}}[|\psi|^{2}]^{\frac{1}{2}}}{M_{\frac{kq}{2}}[|\psi|^{2}]}\,\, E_{\frac{k}{2}}\left[\left\Vert x\right\Vert _{2}^{\gamma}\right]^{\frac{1}{\gamma}}E\left[\left\Vert \xi\right\Vert _{2}^{\theta}\right]^{\frac{1}{\theta}}\geq\frac{n}{2\pi kq}.\label{eq:FirstInequalityForGeneralEuclideanMoments}
\end{alignat}
For $\gamma=\theta=2,$ the lower bound is attained if and only is
$|\psi|$ is a generalized Gaussian. For $\gamma=\theta=2,$ $q=1,$
this inequality yields a multidimensional version of the Weyl-Heisenberg
uncertainty principle.\textup{ For $\frac{3}{2}-\frac{1}{\beta}>q,$
we also get the inequality} 
\[
\left(E_{\frac{k}{2}}\left[\left\Vert x\right\Vert _{2}^{\gamma}\right]\right)^{\frac{1}{\gamma}}\left(E\left[\left\Vert \xi\right\Vert _{2}^{\theta}\right]\right)^{\mbox{\ensuremath{\frac{1}{\theta\lambda}}}}>\frac{1}{M_{\frac{k}{2}}[|\psi|^{2}]^{\frac{1}{k\lambda}}}\,\left(E_{\frac{k}{2}}\left[\left\Vert x\right\Vert _{2}^{\gamma}\right]\right)^{\frac{1}{\gamma}}\left(E\left[\left\Vert \xi\right\Vert _{2}^{\theta}\right]\right)^{\mbox{\ensuremath{\frac{1}{\theta\lambda}}}}\geq K.
\]

\end{proposition}
 \enlargethispage{0.15cm}

\vspace{-0.6cm}

\begin{small}

\end{small}

\end{document}